# Singularity displacement in random speckle patterns of diffusive and localized waves: universality lost and regained

Sheng Zhang[1], Bing Hu[1], Patrick Sebbah[2], and Azriel Z Genack[1]

[1]Department of Physics, Queens College, The City University of New York, Flushing, NY 11367, USA

[2]Laboratoire de Physique de la Matière Condensée/CNRS, Université de Nice-Sophia Antipolis, Parc Valrose 06108 Nice Cedex 02, France

All random wave fields possess a network of phase singularities. We show that while the phase statistics within speckle patterns is generic, the statistics of the motion of phase singularities differs substantially for diffusive and localized waves. This reflects the changing wave interaction with the underlying modes of multiply-scattering systems. Universality is regained when the motion of phase singularities is charted against the phase excursion which reflects the variation of phase change across the speckle pattern. The evolution of speckle patterns can therefore be used to monitor internal motion in complex systems and to characterize the nature of wave propagation.

The superposition of randomly phased waves scattered from a disordered region gives rise to a speckled intensity pattern and a random probability distribution of phase.[1] Nevertheless, the pattern of phase variation is not devoid of structure. A network of vortices of phase circulation centered upon points of vanishing intensity appears.[2,3,4,5,6] The phase jumps discontinuously by π radians along any line passing through these points.[2] The evolution of the speckle pattern is tied to the motion of such phase singularities. Recently, Berry and Dennis[5] calculated the statistics of the velocity of singularities for ergodic wave fields produced by the superposition of randomly phased plane waves. Ergodicity is increasingly violated, however, in the Anderson transition from extended diffusing waves to exponentially peaked localized waves[7,8,9] as a result of enhanced intensity fluctuations[10,11,12,13,14,15]. The question arises as to whether the statistics of the speckle pattern and its evolution are universal or rather reflects the changing character of radiation in the Anderson transition.

In this Letter, we report that though the statistics within single speckle patterns is generic, the statistics of the displacement of phase singularities induced by a frequency shift of the incident radiation is non-universal, differing substantially for diffusive and localized waves. The evolution of the speckle pattern thus provides a diagnostic of the nature of wave propagation. We show, however, that the motion of singularities is universal when charted against the "phase excursion," which is the average deviation of the phase shift from its average value over the speckle pattern. Statistical changes in the speckle pattern are related to changes interaction with electromagnetic modes of the random medium.

The nature of wave propagation is determined by the closeness to the localization threshold. Localization is intimately connected to the enhancements of fluctuations and their correlation with varying voltage[10], sample structure[16], frequency[17], scattering angle[11,18] and position[19] of transmission quantities such as the intensity, total transmission and conductance over the prediction for Gaussian fields. Photon localization can be characterized by the variance of the total transmission coefficient for an incident wave in mode or position, $a$, normalized by its average value over a collection of random sample configurations, $\text{var}(s_a = T_a / \langle T_a \rangle)$.[14] The localization threshold which corresponds to the point at which the dimensionless ratio of the width of modes to their spacing is unity[8] occurs at $\text{var}(s_a) = 2/3$.[14] Measurement of speckle are described below for diffusive and localized waves, with $\text{var}(s_a) = 0.14$ and 3.0, respectively.

The electromagnetic field transmitted through the scattering medium contained in a cylindrical copper tube is picked up by a short wire antenna. The amplitude $|E|$ and phase shift $\varphi$ of the field polarized along the antenna relative to the incident field is detected with use of a vector network analyzer. The sample is composed of alumina spheres with diameter 0.95 cm and refractive index 3.14 embedded in Styrofoam shells of refractive index 1.04 to produce an alumina volume fraction of 0.068. Photons are localized in a narrow window above the first Mie sphere resonance.[20] Microwave spectra are taken on a 1-mm-square grid over the output face. Speckle patterns were obtained for diffusive waves from 14.7 to 15.7 GHz [Fig. 1(a)] and for localized waves from 10 to 10.24 GHz in 40 random configurations.[20] Spectra were taken with frequency steps of 625 kHz for diffusive waves and 300 kHz for localized waves, corresponding to approximately 1/7 of the field correlation frequency in each case. New realizations of the random sample were created after the full speckle pattern is recorded by rotating and vibrating the sample tube.

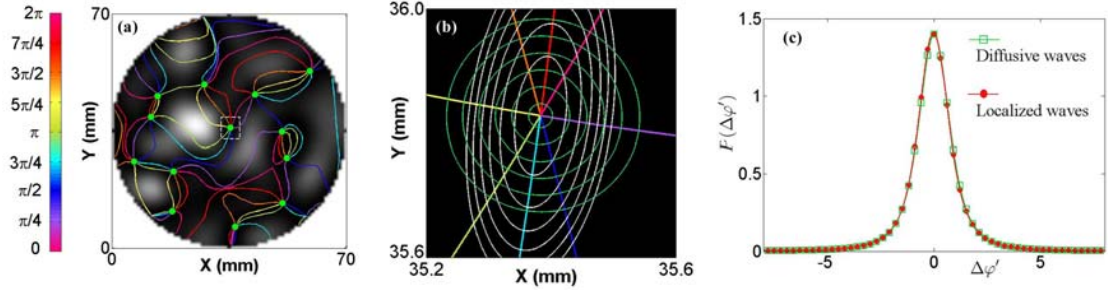

FIG. 1. (a) An example of speckle patterns for diffusive waves. The intensities at measured points are exhibited via a gray scale shown with white corresponding to the maximum and black to zero intensity. Equiphase lines with values of phase, $\varphi = n\pi/4$ rad, with n = 0…7, are shown as different colors indicated in the color bar. Phase singularity points are marked with light green dots. (b) Elliptical contours of intensity (white) and circular contours of current magnitude (green) are found near the singularity within the rectangle in (a). (c) The probability distribution $P(\Delta\varphi')$ for diffusive and localized waves.

Since the measured signal has limited bandwidth in $k$-space, the full speckle pattern can be reconstructed with arbitrary resolution by interpolating between the measured points using the Whittaker-Shannon sampling theorem. Applying this theorem reveals the predicted structures[5] of the diverging phase gradients, elliptical contours of intensity, $I = |E|^2$, and circular contours of current magnitude, $J = I|\nabla\varphi|$ near singularities, which has also been observed in surface scattered optical waves.[6] A magnified view of a typical singularity is shown in Fig. 1(b). Equiphase lines are seen to emanate from phase singularities at points of vanishing intensity. At these points, the phase angle is indeterminate since both the in and out-of-phase components vanish. Equiphase lines with the same slope on different sides of a singularity are seen in Fig. 1(b) to differ in phase by π radians.

To test whether the statistics *within* speckle patterns are generic, we compare the probability distributions, $P(\Delta\varphi' = \Delta\varphi/\overline{|\Delta\varphi|})$, for diffusive and localized waves, of the normalized phase change in a step, $\Delta x = 1$ mm, for which the average magnitude of the phase change within a speckle pattern is small, $\overline{|\Delta\varphi|} \ll 1$. The steps here are taken in a direction perpendicular to the antenna orientation. The distributions shown in Fig. 1(c) are seen to be identical, demonstrating that field statistics within single configurations are universal.

Next we consider whether changes *between* speckle patterns are similarly generic. We analyze the displacement of singularities in the speckle pattern for diffusive and localized waves for a single frequency step. In order to compare the statistics of displacement in different spectral regions, we normalize the displacements by an internal measure of the scale of the speckle pattern, $s = \Delta x/\langle\overline{|\Delta\varphi|}\rangle$, where $\langle...\rangle$ denotes the average over the random ensemble. A measure of the overall variation of the speckle pattern, is provided by the average magnitude of the appropriately normalized displacements of the singularities within each pattern, $\overline{\delta r'} = \overline{\delta r}/s$. The probability distributions of $\overline{\delta r'}$, $P(\overline{\delta r'})$, for diffusive and localized waves are seen in Fig. 2(a) to differ substantially. The probabilities of small and large values of $\overline{\delta r'}$ are higher for localized than for diffusive waves. This difference can be seen directly in the movie clips of the development of speckle patterns with frequency shift,[21] in which the motion of singularities is relatively smooth for diffusive and abrupt for localized waves.

The nonuniversal distribution of $\overline{\delta r'}$ may reflect the distinction that can be drawn between on and off-resonance excitation of electromagnetic modes in localized samples since the linewidth of electromagnetic modes is generally narrower than the spacing between modes as seen in distinct transmission peaks for localized waves [Fig. 3(b)]. In contrast, the linewidth exceeds the spacing between modes for diffusive waves so that the wave is always within the linewidth of several modes [Fig. 3(a)]. Such a distinction is responsible for the difference in the distributions of the phase change of a wave incident on the sample at point $a$ and emerging at point $b$, $\Delta\varphi_{ab}$ for diffusive and localized waves.[22] For a fixed small frequency shift, the phase shift is proportional to the average transit time of photons in a narrow bandwidth pulse, $\tau_{ab} = d\varphi_{ab}/d\omega$, where $\omega = 2\pi\nu$ is angular frequency.[20] The single channel delay time $\tau_{ab}$ and hence the phase shift is larger on resonance than off.[22] The average transit time of transmitted photons is therefore essentially proportional to the average phase change over the whole output surface, $\Delta\varphi_a \equiv \overline{\Delta\varphi_{ab}}$, which we call the "phase advance." We expect that the spatially averaged transit time is longer on resonance than off. This is confirmed by the correspondence of peaks in the phase advance with resonances in total transmission for localized waves seen in Fig. 3(b). If the sensitivity to changes in sample configuration were related to the transit time, we might expect that the conditional probability distribution for given transit time or phase advance, $P_{\Delta\varphi_a}(\overline{\delta r'})$, would be universal. Instead, this distribution is seen in Fig. 2(b) to be different for diffusive and localized waves.

The absence of direct correlation between $\overline{\delta r'}$ and $\Delta\varphi_a$ can be understood by considering the extreme case in which the phase change at each point is a constant, i.e. $\Delta\varphi_a$. In this case, the point of intersection of equiphase lines and hence the location of the singularity does not change despite the constant change in the value of each equiphase line, The point of intersection of equiphases curves only changes when the phase change across the speckle pattern is nonuniform. We therefore consider a measure of phase change which reflects its spatial variation and neglects the average phase shift; the "phase excursion," $\Delta\Phi_a \equiv \overline{|\Delta\varphi_{ab} - \Delta\varphi_a|}$. We find that when the probability distributions of $\overline{\delta r'}$ are plotted for given values of phase excursion, they are the same for diffusive and localized waves [Fig. 2(c)]. This indicates that $P_{\Delta\Phi_a}(\overline{\delta r'})$ is universal.

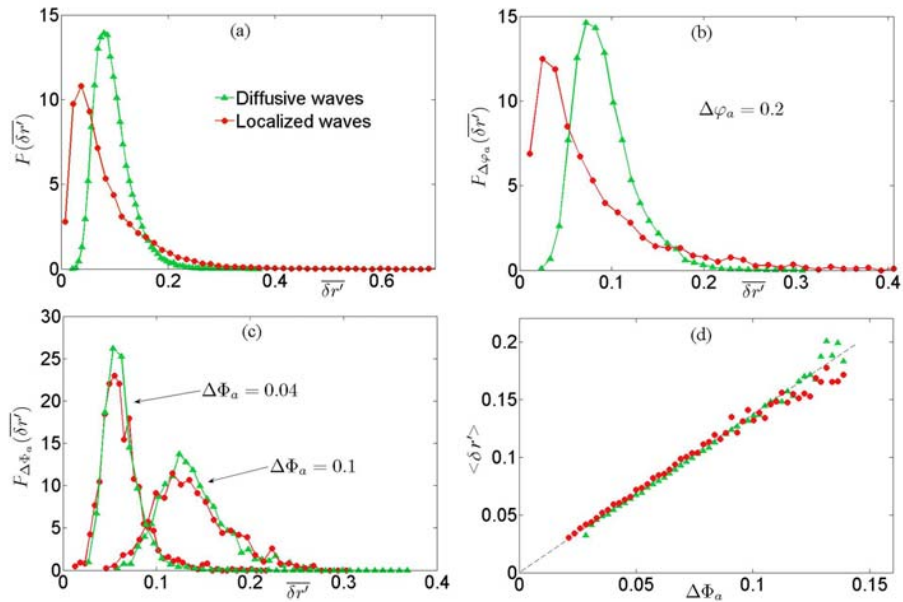

FIG. 2. Green triangles are for diffusive waves and red circles are for localized waves. (a) Probability distributions of normalized average magnitude of displacement, $\overline{\delta r'}$ for fixed frequency shift. (b) Conditional probability distributions of $\overline{\delta r'}$ for fixed phase advance of $\Delta\varphi_a = 0.2$ rad between spectra. (c) Conditional probability distribution of $\overline{\delta r'}$ for values of phase excursions, $\Delta\Phi_a = 0.04$ and $0.1$ rad. (d) Relation between average singularity displacement and phase excursion.

The universality of statistical changes in the speckle pattern for given phase excursion is also seen in the overlap of the curves of the ensemble average displacement $\langle \delta r' \rangle$ versus $\Delta\Phi_a$ for a single frequency step for diffusive and localized waves [Fig. 2(d)]. A linear fit of the variation of $\langle \delta r' \rangle$ with $\Delta\Phi_a$ which passes through the origin for values of $\Delta\Phi_a <$ 0.1 gives the slope, $\eta = \langle \delta r' \rangle / \Delta\Phi_a = 1.375$. We expect that for small changes in phase excursion, $\eta$ is a universal constant of transformation of the speckle pattern in random ensembles, being independent of the cause of the change in the speckle pattern, whether it is internal rearrangement of the sample, width of the spectrum of the incident wave, variation of distance of source or detector from the sample, time delay from a pulsed source or frequency shift of a monochromatic incident wave. The universality of $P_{\Delta\Phi_a}(\overline{\delta r'})$ and the linear relation between $\langle \delta r' \rangle$ with $\Delta\Phi_a$ show that both the average singularity displacement and the phase excursion are fundamental measures of the degree of change of a speckle pattern. However, since all measured points are involved in calculating the phase excursion as compared to the much smaller number of singularities, fluctuations are smaller in $\Delta\Phi_a$ than in $\overline{\delta r'}$. The spectra of $\Delta\Phi_a$ are consequently relatively smooth as compared to spectra of $\overline{\delta r'}$.

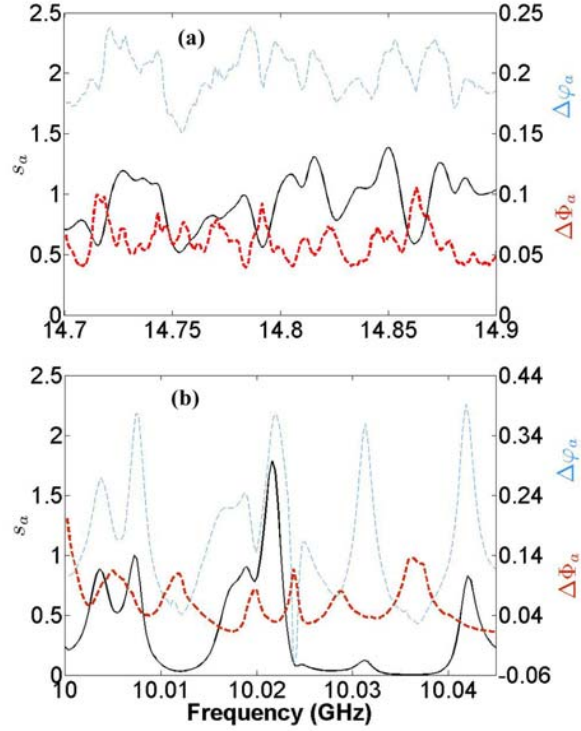

**FIG. 3. Examples of spectra of phase excursion $\Delta\Phi_a$ (red thick dashed lines), phase advance $\Delta\varphi_a$ (blue thin dashed lines) and normalized total transmission $s_a$ (black solid lines) for (a) diffusive and (b) localized wave.**

The reason that the displacement of singularities is related to the phase excursion but not the phase advance can be found by comparing typical spectra of these quantities with spectra of total transmission (Fig. 3). Fluctuations of all quantities are relatively small for diffusive waves because the incident wave always falls within the linewidth of several electromagnetic modes. In contrast, sharp fluctuations are observed in spectra for localized waves since at some frequencies transmission may be dominated by the resonant interaction with a single electromagnetic mode while at others the wave may not be resonant with any mode. The phase advance is large when the wave is tuned through long-lived resonant modes while the phase excursion is generally small through the resonance and peaks between transmission resonances (Fig. 3). This suggests that the spatial distribution of the field does not vary appreciably as long as a single mode dominates transmission. However, the spatial distribution of the wave can be expected to change in the wings of a transmission line at the point that the frequency is tuned sufficiently off resonance that the contributions of other modes influence the speckle pattern. The speckle pattern then changes rapidly leading to rapid motion of the singularities accompanied by a large variation of the phase across the speckle pattern.

We have found that the statistics of two different quantities, one a measure of phase change and the other a measure of displacements of nulls in the intensity pattern, are sensitive to the degree to which modes can be differentiated. Thus we expect that the variances of $\Delta\Phi_a$ or $\overline{\delta r'}$ provide measures of the closeness to the localization threshold along with traditional measures of the scaling of transmission[17,23], transmission fluctuation[14,15], long range correlation[19] and coherent backscattering.[24] In the diffusive limit, changes in the speckle pattern as measured by the phase excursion or the displacement of singularities tend to be uniform with frequency shift. We expect that these parameters of the speckle pattern will similarly reflect the internal rearrangement of structures illuminated with monochromatic radiation. Such measurements may therefore be used to characterize changes in a variety of systems including geological formations, ice caps, and aerospace components, and might be exploited to monitor the heart undergoing cardiac fibrillation.[25]

We thank A. A. Chabanov for discussions of localization. This work was supported by the NSF under grant number DMR-0538350.